\documentclass[runningheads]{llncs}
\usepackage[utf8]{inputenc}

\pdfoutput = 1
\usepackage{amsfonts, amsmath, mathtools, enumerate, amssymb, hyperref, graphics, enumerate, dsfont, csquotes, amsthm, paralist, bbm, url}
\usepackage[english]{babel}
\usepackage[nice]{nicefrac}

\usepackage[sorting=ynt, url = false, doi = false, isbn =false, backend = biber, style = lncs]{biblatex}
\usepackage{textcomp}

\title{The Secretary Problem with Distributions}
\author{Pranav Nuti\inst{}}
\institute{Stanford University, Stanford CA 94305, USA 
\email{pranavn@stanford.edu}\\}

\begin{document}

\maketitle

\begin{abstract}
In the secretary problem, we are presented with $n$ numbers, adversarially chosen and then randomly ordered, in a sequential fashion. After each observation, we have to make an irrevocable decision about whether we would like to accept a number or not, with the goal of picking the largest number with the highest probability. The secretary problem is a fundamental online selection problem with a long history.

A natural variant of the problem is to assume that the $n$ numbers come from $n$ independent distributions known to us. Esfandiari et al. [AISTATS 2020] studied this problem, found the optimal probability under a restrictive ``no-superstars" assumption on the distributions, and conjectured that this assumption could be dropped. In this paper, we prove that this is indeed the case while significantly simplifying both the optimal algorithm and its analysis. We then extend this result in two directions. First, we are able to relax the assumption of independence to a kind of negative dependence, demonstrating the robustness of our algorithm. Second, we are able to replace knowledge of the distributions with a (more realistic) knowledge of samples from the distributions.
\end{abstract}

\section{Introduction}

In the classical secretary problem, we are presented with $n$ numbers, adversarially chosen and then randomly ordered, in a sequential fashion. After each observation, we have to make an irrevocable decision about whether we would like to accept a number or not, with the goal of picking the largest number with the highest probability. It is well known that this goal can be achieved with a probability of $\frac{1}{e}$.

 In this paper, we are interested in how access to more information about how the $n$ numbers are chosen affects the optimal probability of success. Most relevant to our results is the work of Gilbert and Mosteller \cite{GilbertMosteller} who studied the case where the $n$ numbers, instead of being adversarially chosen, are $n$ independent draws from a \textit{single} distribution known to us, and found the optimal probability of success to be $\gamma \approx0.5801$.

A natural extension of the work of Gilbert and Mosteller is to assume that the $n$ numbers come from $n$ \textit{different} independent distributions which are known to us. As in the classical secretary problem, the numbers are randomly ordered before we observe them. 

Esfandiari et al. \cite{Superstars} studied this problem (which they call the ``best-choice prophet secretary problem") under a restrictive ``no-superstars" assumption on the distributions and showed that the optimal probability of success is $\gamma \approx0.5801$, the same as in the work of Gilbert and Mosteller. Distributions are said to satisfy a no-superstars assumption if none of the distributions are (a priori) particularly likely to yield the largest draw. Esfandiari et al.  conjectured that this assumption could be dropped, and asked for a simpler proof of their result. 

\subsection{Our contributions} In this paper, we prove the conjecture of Esfandiari et al. \cite{Superstars} and significantly simplify the optimal algorithm and its analysis.

Our proof shows that this optimal probability $\gamma$ can be attained by using ``threshold-based" algorithms, i.e., algorithms of the following form: accept the $i^\text{th}$ number if it is the largest seen so far, and is greater than a threshold $\tau_i$. The most important piece of our analysis is a lemma regarding the distribution of the maximum of $n$ random variables which plays a vital role in allowing us to generalize the work of Gilbert and Mosteller.

We then investigate our result in two further directions. First, we are able to relax the assumption of independence to a kind of negative dependence, demonstrating the robustness of our algorithm. A crucial tool in this analysis is the use of Han's inequality for submodular set functions. Our method of applying Han's inequality together with this particular negative dependence criterion is flexible, and can also be used to generalize other results about secretary problems beyond the case of independent distributions. The number of balls in each bin when $m$ balls are dropped randomly and independently into $n$ bins satisfy our negative dependence criterion. 

Second, we are able to replace knowledge of the distributions with a (more realistic) knowledge of samples from the distributions. Specifically, we are able to show that given enough samples from each  distribution, we can approximate the thresholds in our optimal algorithm so well that we can succeed with a probability as close to $\gamma \approx0.5801$ as we please. It is critical to note that the number of samples needed from each distribution \textit{does not} depend on $n$. 

\subsection{Related literature} Over the years, many variations on the secretary problem in which the numbers come from independent distributions have been explored. We discuss here some work different from our own along four particular dimensions.

The first dimension concerns the objective. In the secretary problem with distributions (or the ``best-choice prophet secretary problem"), the goal is to maximize the probability of picking the largest number (the ``secretary objective"). Alternatively, one's goal might be to maximize the expected value of the picked number. This problem was introduced in \cite{prophetsecretary} by Esfandiari et al., and their results were improved in \cite{beatinge}. The optimal answer remains unknown, but the work of Correa et al. \cite{blindstrategies} provides the best known bounds.

The second dimension considers whether the distributions are identical or not. In this paper, we show that the distributions being non-identical does not decrease the probability of success; it remains the same as the IID (identical distribution) setting of Gilbert and Mosteller. However, the situation with the expected value objective is different, and the distributions being non-identical \textit{does} decrease the optimal reward (see \cite{iidprophet} and \cite{blindstrategies} for further discussion).

The third dimension considers the order in which the distributions are observed. In our paper, the numbers are observed in random order. Instead, the distributions may be adversarially ordered. When considered with the expected value objective, this is the the classical prophet inequality  (see \cite{prophetsurvey} for a recent survey). The same problem with the secretary objective is addressed in \cite{Superstars}. Recent work has also considered the possibility of other orderings (the so-called ``constrained-order" prophet inequalities \cite{constrained}).

The fourth dimension concerns how much information about the distributions is known. Some recent papers have focused on  exploring which of the results from the full-knowledge-of-distributions setting discussed above can be replicated under access to samples from the distributions (see, for example, \cite{sampledriven}, \cite{googol}, \cite{rubinstein}, \cite{scoopy}). The most surprising of these results is perhaps the work of Rubinstein et al. \cite{rubinstein}, which demonstrates that the reward of the classical prophet inequality can be obtained with access to just a \textit{single} sample from each distribution. 

\subsection{Organization} In section 2, we present a statement and proof of the conjecture of Esfandiari et al. In section 3, we discuss how the independence assumption can be relaxed to a kind of negative dependence. In section 4, we describe how knowledge of samples from the distributions suffices to obtain a success probability arbitrarily close to $\gamma \approx0.5801$.

\section{The secretary problem with distributions}

An adversary chooses independent continuous distributions $\mathcal{D}_1, \ldots, \mathcal{D}_n$ and generates draws $X_1, \ldots, X_n$ from these distributions. We are told what the distributions $\mathcal{D}_1, \ldots, \mathcal{D}_n$ are. Next, the draws are randomly ordered and we are sequentially presented with them. We have to make an irrevocable decision about whether we would like to accept a draw or not, with the goal of selecting the largest number. What is the probability with which we can succeed?

\begin{theorem}
The probability of success in the secretary problem with distributions is at least $\gamma \approx 0.5801$, the probability of success when the distributions are all identical.
\end{theorem}

\begin{proof}

Fix parameters $\{d_i\}_{i=1}^n$ (to be picked later) and let us choose $\tau_i$, which will be our thresholds, such that $\Pr[\max_{k=1}^n X_k \leq \tau_i] = d_i^n$ (all superscripts are exponents). Assume these parameters are monotonically decreasing in $i$. (If the $X_i$ were IID, and uniform on $[0,1]$, then $\tau_i = d_i$.)

We have the following simple, but vital lemma:

\begin{lemma} Suppose that $X_1, X_2,\ldots, X_n$ are independent.  If $r$ of the $X_k$ are randomly chosen, say, $X_{j_1}, X_{j_2}, \cdots, X_{j_r}$, then, $\Pr[\max_{k=1}^r X_{j_k} \leq \tau_i] \geq d_i^r$.
\end{lemma}

\begin{proof} Fix a constant $T$, and suppose $\Pr[X_k \leq T] = a_k$. Then, $$\Pr[\max_{k=1}^n X_k \leq T] = \prod_{k=1}^n a_k$$  If $r$ of the $X_k$ are randomly chosen, say, $X_{j_1}, X_{j_2}, \cdots, X_{j_r}$  then $$\Pr[\max_{k=1}^r X_{j_k} \leq T] = \frac{\sum_{(i_1, i_2, \ldots, i_r)}\prod_{k=1}^r a_{i_k}}{\binom{n}{r}}$$

By the AM-GM inequality (which tells us that the arithmetic mean of a set of numbers is at least as large as its geometric mean), this is at least 

$$\left(\prod_{k= 1}^n a_k^{\binom{n-1}{r-1}}\right)^{\frac{1}{\binom{n}{r}}} = \left(\prod_{k=1}^n a_k\right)^{\frac{r}{n}} $$

In particular, we can let $T = \tau_i$, and the desired result follows.  
\end{proof}

Shortly, we shall be applying the lemma to the situation where $X_{j_1}, X_{j_2}, \cdots, X_{j_r}$ are the first $r$ draws we are presented with (which indeed constitute a randomly chosen subset of the $X_i$ of size $r$).

Consider a strategy in which we accept the $i^\text{th}$ number if and only if it is the largest seen so far, and is more than its threshold $\tau_i$  (Correa et al. \cite{blindstrategies} call such strategies $\textit{blind}$). We analyze the probability of succeeding with this strategy, following the proof of Gilbert and Mosteller \cite{GilbertMosteller}.

Note that for any $i \leq r$, the probability that the $i^\text{th}$ number is largest amongst the first $r$, but is less than its threshold is just

$$\frac{\Pr[\max_{k=1}^r X_{j_k} \leq \tau_i]}{r}\geq \frac{d_i^r}{r}$$since this happens if (and only if!) all the first $r$ numbers are less than $\tau_i$ and they're ordered so that the $i^\text{th}$ number is largest.

Next, note that the probability the $i^\text{th}$ number is less than its threshold, is largest amongst the first $r$, but is not largest amongst all numbers is at least 

$$\frac{d_i^r}{r}-\frac{\Pr[\max_{k=1}^n X_k \leq \tau_i]}{n} = \frac{d_i^r}{r} - \frac{d_i^n}{n}$$since the second term simply represents the probability that the $i^\text{th}$ number is  largest amongst the first $n$ and is less than its threshold.

Now, note that the probability that the $i^\text{th}$ number is less than its threshold, is  largest amongst the first $r$, but the $r+1^{\text{th}}$ number is  largest amongst all numbers is at least 

$$\frac{\frac{d_i^r}{r}-\frac{d_i^n}{n}}{n-r}$$since it's equally likely that each of the last $n-r$ numbers turns out to be the largest amongst them.

Furthermore, note that the above is also actually the probability \textit{no number before the $r+1^{\text{th}}$ is  chosen}, the $i^\text{th}$ number is  largest amongst the first $r$, but the $r+1^{\text{th}}$ number is  largest amongst all numbers. This is because if the largest number amongst the first $r$ is less than its threshold, it cannot be chosen; certainly, no number after it (amongst the first $r$) can be chosen; no number before it can be chosen either since it wouldn't have been able to meet its threshold (since the $\tau_i$ are decreasing).

Thus we conclude that the the probability no number before the $r+1^{\text{th}}$ is  chosen, and the $r+1^{\text{th}}$ number is largest amongst all numbers is at least

$$\sum_{i = 1}^r\frac{\frac{d_i^r}{r}-\frac{d_i^n}{n}}{n-r}$$

The probability that the  $r+1^{\text{th}}$ number is largest amongst all numbers but is less than its threshold is just

$$\frac{\Pr[\max_{k=1}^n X_k \leq \tau_{r+1}]}{n} = \frac{d_{r+1}^n}{n}$$

We conclude, that the probability of succeeding by picking the $r+1^{\text{th}}$ number is at least

$$\left(\sum_{i = 1}^r\frac{\frac{d_i^r}{r}-\frac{d_i^n}{n}}{n-r}\right)-\frac{d_{r+1}^n}{n}$$

The probability of succeeding by picking the first number is 

$$\frac{1 - d_1^n}{n}$$

The overall probability of succeeding is the sum of the probabilities of succeeding by picking any particular number, so our arguments thus far demonstrate that the probability of success for our threshold-based algorithm is at least

$$\frac{1 - d_1^n}{n} + \sum_{r = 1}^{n-1}\left( \left(\sum_{i = 1}^r\frac{\frac{d_i^r}{r}-\frac{d_i^n}{n}}{n-r}\right)-\frac{d_{r+1}^n}{n}\right)$$

In the IID setting, we actually have $\Pr[\max_{k=1}^r X_{j_k} \leq \tau_i] = d_i^r$. Therefore, the above formula is in fact exact in the IID setting! This demonstrates that for any threshold-based algorithm in the IID setting, there is a threshold-based algorithm in our more general setting that does at least as well.

Since it is known that a threshold algorithm of this sort is in fact optimal when the distributions are identical (this claim is proved in section 2 of \cite{formalsecretary}), we conclude that we can always do at least as well as we do in the IID setting and to achieve a probability of success of at least $\gamma$, we just need to choose the $d_i$ which are optimal when the distributions are identical.

Note of course that it is impossible to do better than $\gamma$ in general, because $\gamma$ is the optimal answer when the distributions are identical.

We now briefly discuss how the optimal probability of success is calculated in the IID setting (see \cite{formalsecretary} for details of these calculations).

Our computation above immediately yields that the optimal probability of success in the IID setting is decreasing in $n$, since if $Z$ is a random variable deterministically equal to 0 and $X_1, X_2, \ldots, X_{n+1}$ are IID uniform $[0,1]$, then 
\begin{align*}
    &\text{   Pr}[\text{succeeding with draws }X_1, X_2, \ldots, X_n]\\ 
    = &\text{   Pr}[\text{succeeding with draws }X_1, X_2, \ldots, X_n, Z]\\ 
    \geq & \text{   Pr}[\text{succeeding with draws }X_1, X_2, \ldots, X_{n+1}]
\end{align*}where the equality in the first line follows from the fact that the optimal algorithm with draws $X_1, X_2, \ldots, X_n, Z$ is to just act as if the $Z$ never appeared, and the inequality from the second line is just our claim that we can always do at least as well in the IID setting.

In particular, this means that the worst case for the IID setting arises when we take the limit as $n \to \infty$. (This is proved slightly differently in section 3 of \cite{formalsecretary}.) 

The optimal value of $d_{n-i}$ turns out to not depend on $n$. Explicit expressions for the optimal values of $d_{n-i}$ can be found ($d_{n-i} = 1 - \frac{c}{i} + O\left(\frac{1}{i^2}\right), \int_{0}^{c}\frac{e^x - 1}{x}dx = 1$) \cite{GilbertMosteller}. 

We can then evaluate the success probability in the limit as $n \to \infty$ by plugging in the optimal value of $d_i$ into our formula for the success probability of a threshold algorithm. The success probability turns out to equal $(e^c - c - 1)\int_c^{\infty}\frac{e^{-x}}{x}dx + e^{-c} \approx 0.5801$ (see section 4 of \cite{formalsecretary}).  
\end{proof}

We should note that the assumption that the distributions are continuous was not important. We can achieve the same result when the distributions are possibly discrete. To do this, we employ the same trick used 
in \cite{rubinstein}. Replace $\mathcal{D}_i$ with a bivariate distribution, where the first coordinate is from $\mathcal{D}_i$ and the second coordinate is a value drawn independently and uniformly from $[0, 1]$. Impose the lexicographic order on $\mathbb{R}^2$, and then, there is no difficulty in determining the value of $\tau_i$ from $d_i$. While implementing the algorithm, we can generate a random value from $[0,1]$ for every draw presented to us to decide whether it is large enough to accept, and the guarantee on the success probability remains the same.

\section{Extension to negative dependence}

In this section, we shall drop the assumption that the draws, $X_i$, are independent. Let us write $Y_i = \mathbbm{1}_{X_i \leq T}$, and assume instead that the $Y_i$ are \textit{conditionally negatively associated} for every $T$. What does this mean?

We say that random variables $Y_1, Y_2, \ldots, Y_n$ are \textit{negatively associated} if for any non-intersecting $I, J \subset [n]$, and any pair of non-decreasing functions $f$ and $g$, 

$$\mathbb{E}[f(Y_I)g(Y_J)]\leq \mathbb{E}[f(Y_I)]\mathbb{E}[g(Y_J)]$$

Furthermore, the random variables $Y_1, Y_2, \ldots, Y_n$ are said to be \textit{conditionally negatively associated} if they remain negatively associated even after conditioning upon the values of some subset of the random variables $Y_1, Y_2, \ldots, Y_n$.

A full discussion of this particular negative dependence criterion is outside the scope of this paper (see \cite{pemantle} for an introduction to various negative dependence criteria). One reason to consider this criterion is that there is a very natural example of distributions that satisfy it: Suppose $X_i$ is a draw from distribution $\mathcal{D}_i$, the number of balls in the $i^\text{th}$ bin when $m$ balls are dropped randomly and independently into $n$ bins. Then, the $Y_i$ are conditionally negatively associated \cite{ballsandbins}.

The following result recovers the lemma from section 2 under the assumption that the $Y_i$ are conditionally negatively associated:

\begin{lemma}
Suppose that $Y_i = \mathbbm{1}_{X_i \leq T}$, and the $Y_i$ are conditionally negatively associated.  If $r$ of the $X_k$ are randomly chosen, say, $X_{j_1}, X_{j_2}, \cdots, X_{j_r}$, then, $\Pr[\max_{k=1}^r X_{j_k} \leq T] \geq \Pr[\max_{k=1}^n X_{k} \leq T] ^{\frac{r}{n}}$.
\end{lemma} 

\begin{proof} Since the $Y_i$ are conditionally negatively associated, we know that after conditioning on $S = \{\prod_{i \in A \cap B}Y_i = 1\}$, we have $$\mathbb{E}\left[\prod_{i \in A \backslash B}Y_i\Big |S\right]\cdot \mathbb{E}\left[\prod_{i \in B \backslash A}Y_i\Big |S\right] \geq \mathbb{E}\left[\prod_{i \in A \backslash B \cup B \backslash A}Y_i\Big |S\right]$$ which implies (with no conditioning!) that$$\frac{\mathbb{E}\left[\prod_{i \in (A \backslash B)\cup (A \cap B)}Y_i\right]}{\mathbb{E}\left[\prod_{i \in A \cap B }Y_i\right]}\cdot \frac{\mathbb{E}\left[\prod_{i \in (B \backslash A)\cup (A \cap B)}Y_i\right]}{\mathbb{E}\left[\prod_{i \in A \cap B }Y_i\right]}  \geq \frac{\mathbb{E}\left[\prod_{i \in A \backslash B \cup B \backslash A\cup (A \cap B)}Y_i\right]}{\mathbb{E}\left[\prod_{i \in A \cap B }Y_i\right]} $$ or in other words,$$\mathbb{E}\left[\prod_{i \in A}Y_i\right]\cdot \mathbb{E}\left[\prod_{i \in B }Y_i\right] \geq \mathbb{E}\left[\prod_{i \in A \cup B }Y_i\right]\cdot \mathbb{E}\left[\prod_{i \in A \cap B }Y_i\right]$$ 

So it follows that if we define $g(A) = \log \mathbb{E}\left[\prod_{i \in A}Y_i\right] = \log \Pr[\max_{k\in A} X_{k} \leq T]$, then $g$ is a submodular set function.

Therefore, it follows by Han's inequality for submodular set functions (see pages 16--18 in \cite{hans} for a statement and proof of Han's inequality), 

$$\frac{r}{n}\cdot g([n]) \leq \frac{\sum_{|A|=r}g(A)}{\binom{n}{r}}$$

Hence, we must have

$$\frac{r}{n}\cdot \log \mathbb{E}\left[\prod_{i}Y_i\right] \leq \frac{\sum_{|A|=r}\log \mathbb{E}\left[\prod_{i \in A}Y_i\right]}{\binom{n}{r}}$$

This means that

$$\Pr[\max_{k=1}^n X_{k} \leq T] ^{\frac{r}{n}} \leq {\prod_{|A|=r}\mathbb{E}\left[\prod_{i \in A}Y_i\right]}^{\frac{1}{\binom{n}{r}}}$$

By the AM-GM inequality, we know that

$$ {\prod_{|A|=r}\mathbb{E}\left[\prod_{i \in A}Y_i\right]}^{\frac{1}{\binom{n}{r}}} \leq \frac{\sum_{|A|=r}\mathbb{E}\left[\prod_{i \in A}Y_i\right]}{\binom{n}{r}}$$

But, if $r$ of the $X_k$ are randomly chosen, say, $X_{j_1}, X_{j_2}, \cdots, X_{j_r}$  then $$\Pr[\max_{k=1}^r X_{j_k} \leq T] = \frac{\sum_{|A|=r}\mathbb{E}\left[\prod_{i \in A}Y_i\right]}{\binom{n}{r}}$$

The desired result follows.  
\end{proof}

Now, the only place in which the results of section 2 depended on the independence of the $X_i$ was in the proof of the lemma. It follows that the results of section 2 also apply to the $X_i$ when the $Y_i$ are conditionally negatively associated. 

In particular, this means, in the secretary problem with balls and bins, in which we observe a sequence of $n$ bins (into which $m$ balls have been randomly and independently dropped) and we have to select the bin with the most balls, we can succeed with a probability of at least $\gamma$. 

Our method of applying Han's inequality together with this particular negative dependence criterion is flexible, and can also be used to generalize other results about secretary problems beyond the case of independent distributions (for example, theorem 1 in \cite{Superstars} which is similar to the question that this paper addresses, but with an adversarial ordering of distributions).

\section{Limited knowledge: samples from distributions}

In this section, we aim to show that with sufficiently many samples, we can find thresholds $T_i$ amongst numbers in the samples which estimate the ``real" thresholds we would've used with full knowledge of the distributions. We have not attempted to optimize any of the arising constants. 

Suppose we have a distribution $\mathcal{D}$, and independent of $\mathcal{D}$, probabilities $p_i$, with $p_i > \delta$. Fix a parameter $\varepsilon<\frac{1}{10}$. Let $X$ be a random variable with distribution $\mathcal{D}$ and suppose we draw $m$ samples independently from $\mathcal{D}$. Essential to the discussion in this section is the following lemma (quite similar to Lemma 8 in \cite{rubinstein}):

\begin{lemma}
If $m$ is sufficiently large (as a function of only $\varepsilon$ and $\delta$), with probability arbitrarily close to 1 we can find numbers $T_i$ amongst the $m$ samples so that $\frac{p_i}{(1+\varepsilon)^2}\leq\Pr[X \leq T_i]\leq p_i(1+\varepsilon)$ for every $i$ simultaneously.

\end{lemma} 

\begin{proof}
We need to find numbers $T_i$ amongst the samples so that $\Pr[X \leq T_i]\approx p_i$. To do this, we only need to find numbers $M_k$ such that $\Pr[X\leq M_k]\approx \frac{1}{(1+\varepsilon)^k}$ for $k$ such that $\frac{1}{(1+\varepsilon)^k} > \frac{\delta}{1+ \varepsilon}$, because then, if $p_i \in \left[\frac{1}{(1+\varepsilon)^{k}}, \frac{1}{(1+\varepsilon)^{k-1}}\right)$, $\Pr[X\leq M_k]$ is also approximately $p_i$, so we can let $T_i=M_k$.
 
Accordingly, let us define for $k \in \mathbb{N}$, $m_k  = \frac{m}{(1+\varepsilon)^{k} }$, $M_k =m_k^{\text{th}}$ smallest sample, and $A_k$ is such that $\Pr[X \leq A_k] = \frac{1}{(1+ \varepsilon)^k}$. We know that the expected number of samples $\leq A_k$ is $m_k$, and the number of samples $\leq A_k$ is a sum of independent Bernoulli random variables, so by the multiplicative Chernoff bound, the probability that the number of samples is not between $m_{k-1}$ and $m_{k+1}$ is bounded by $2\exp(\frac{-\varepsilon^2m_k}{3})$. 

By the union bound, we conclude that the number of samples $\leq A_k$ is between $m_{k-1}$ and $m_{k+1}$ simultaneously for every $k$ such that $\frac{1}{(1+\varepsilon)^k} > \frac{\delta}{(1+ \varepsilon)^2}$ with probability at least $1-2\left(\frac{-\log \delta}{\log(1+\epsilon)}+2\right)\exp(\frac{-\varepsilon^2\delta m}{3(1+\varepsilon)^2})$. Note that if $m$ is sufficiently large, this probability is arbitrarily close to 1.

It follows that $M_{k-1} \geq A_k \geq M_{k+1}$ (for $k = 1, 2,\ldots, \frac{-\log \delta}{\log(1+\epsilon)}+2$) with probability close to 1, and so, we have $A_{k+1}\leq M_k\leq A_{k-1}$ (for $k = 1, 2,\ldots, \frac{-\log \delta}{\log(1+\epsilon)}+1$)  with probability close to 1. Hence, we conclude that $\frac{1}{(1+\varepsilon)^{k+1}}\leq\Pr[X\leq M_k]\leq \frac{1}{(1+\varepsilon)^{k-1}}$ with probability close to 1.

So if $p_i \in \left[\frac{1}{(1+\varepsilon)^{k}}, \frac{1}{(1+\varepsilon)^{k-1}}\right)$, then $\frac{p_i}{(1+\varepsilon)^2}\leq\Pr[X \leq M_k]\leq p_i(1+\varepsilon)$, as required.  
\end{proof}

We can now prove the main theorem of this section:

\begin{theorem}
There is a function $N$ so that with just $N(\epsilon)$ (independent of $n$) samples from each distribution $\mathcal{D}_i$ (but without knowledge of the distributions themselves), we can succeed with a probability of $\gamma - \epsilon$ in the secretary problem with distributions $\mathcal{D}_i$.
\end{theorem}

\begin{proof}
In the following, $f_1(\epsilon), f_2(\epsilon),$ and $f_3(\epsilon)$ are functions which tend to $0$ as $\epsilon \to 0$ which we have not attempted to optimize (but we can think for the purpose of the proof that $f_1(\epsilon) \approx f_2(\epsilon)\approx \frac{\epsilon}{10}, f_3(\epsilon) \approx -\frac{\epsilon}{100 \log \epsilon} $).

Note that  we can safely ignore the last $f_1(\epsilon)$ fraction of draws, and never choose them, and this will effect our success probability by at most $f_1(\epsilon)$. Furthermore, as long as $i \leq (1-f_1(\epsilon))n$, for $d_i =1 - \frac{c}{n - i} + O(\frac{1}{(n - i)^2})$ (the optimal value), $d_i^n$ is at least $\approx e^{\frac{-c}{f_1(\epsilon)}}$.

Applying the lemma while letting $\mathcal{D}$ be the distribution of $\max_{k=1}^n X_k$, $p_i = d_i^n, \delta \approx e^{\frac{-c}{f_1(\epsilon)}}$ and $\varepsilon \approx \frac{f_3(\epsilon)}{4}$, we conclude that we can find $T_i$ such that with probability $1-f_2(\epsilon)$, we have that $\Pr[\max_{k=1}^n X_k \leq T_i]$ is within a factor of $1+ f_3(\epsilon)$ of $d_i^n$ for every $i$ simultaneously (where $X_i$ are the draws we must choose from).

Now, if $\Pr[\max_{k=1}^n X_k \leq T_i]$ is within a factor of $1+ f_3(\epsilon)$ of $d_i^n$, then we must have that the lower bound we have used for $\Pr[\max_{k=1}^r X_{j_k} \leq T_i]$, $\Pr[\max_{k=1}^n X_k \leq T_i]^{\frac{r}{n}}$ is within a factor of $1+ f_3(\epsilon)$ of $d_i^r$ as well (where $\Pr[\max_{k=1}^r X_{j_k} \leq T_i]$ is the probability that $r$ randomly chosen draws are all less than $T_i$). 

From the expression for the success probability derived in section 2 (remembering that we only need to consider this for $r \leq (1-f_1(\epsilon)n)$), we see that using the estimated thresholds (which are good with a probability of $(1 - f_2(\epsilon))$), we succeed with probability $\gamma - \epsilon$, as long as $f_1(\epsilon)$ $f_2(\epsilon)$, and $f_3(\epsilon)$ are sufficiently small (which can all be achieved with sufficiently many samples). 
\end{proof}

\printbibliography

\end{document}